\documentclass[submission,copyright,creativecommons]{eptcs}
\providecommand{\event}{TARK 2021} % Name of the event you are submitting to
\usepackage{breakurl}             % Not needed if you use pdflatex only.
\usepackage{underscore}           % Only needed if you use pdflatex.
\usepackage{comment}
\usepackage{graphicx}
\usepackage{colortbl}
\usepackage{tikz}
\usepackage{pgf}
\usepackage{color}
\usepackage{comment}

\usetikzlibrary{arrows}
\usetikzlibrary{patterns,automata,arrows,shapes,snakes,topaths,trees,backgrounds,positioning,through,calc}

\usepackage{amsmath,amsthm} 
\usepackage{amssymb}
\usepackage{placeins}
\usepackage{afterpage}
\theoremstyle{definition}
\newtheorem{theorem}{Theorem}

\newtheorem{proposition}[theorem]{Proposition}

\newtheorem{definition}[theorem]{Definition}

\newtheorem{lemma}[theorem]{Lemma}

\newtheorem{example}[theorem]{Example}

\newtheorem{remark}[theorem]{Remark}

\title{Measuring Violations of Positive Involvement in Voting\thanks{We thank the three anonymous reviewers for TARK for their helpful comments.}}
\author{Wesley H. Holliday
\institute{University of California, Berkeley}
\email{wesholliday@berkeley.edu}
\and
Eric Pacuit
\institute{University of Maryland}
\email{epacuit@umd.edu}
}
\def\titlerunning{Measuring Violations of Positive Involvement in Voting}
\def\authorrunning{Holliday and Pacuit}
\begin{document}
\maketitle

\begin{abstract} In the context of computational social choice, we study voting methods that assign a set of winners to each profile of voter preferences. A voting method satisfies the property of positive involvement (PI) if for any election in which a candidate $x$ would be among the winners, adding another voter to the election who ranks $x$ first does not cause $x$ to lose. Surprisingly, a number of standard voting methods violate this natural property. In this paper, we investigate different ways of measuring the extent to which a voting method violates PI, using computer simulations. We consider the probability (under different probability models for preferences) of PI violations in randomly drawn profiles vs. profile-coalition pairs (involving coalitions of different sizes).  We argue that in order to choose between a voting method that satisfies PI and one that does not, we should consider the probability of PI violation conditional on the voting methods choosing different winners. We should also relativize the probability of PI violation to what we call voter potency, the probability that a voter causes a candidate to lose. Although absolute frequencies of PI violations may be low, after this conditioning and relativization, we see that under certain voting methods that violate PI, much of a voter's potency is turned against them---in particular, against their desire to see their favorite candidate elected.
\end{abstract}

\section{Introduction}\label{Introduction}

Voting provides a mechanism for resolving conflicts between the preferences of multiple agents in order to arrive at a group choice. Although traditional voting theory is largely motivated by the example of democratic political elections, the field of computational social choice \cite{Brandt2013,Handbook2016,Endriss2017} views voting theory as applicable to many other preference aggregation problems for multiagent systems.

One of the most basic ideas in voting is that an unequivocal increase in support for a candidate should not result in that candidate  going from being a winner to being a loser. There are at least two ways to formalize this idea. First, there is the fixed-electorate axiom of \textit{monotonicity}: if a candidate $x$ is a winner given a preference profile $\mathbf{P}$, and $\mathbf{P}'$ is obtained from $\mathbf{P}$ by one voter moving $x$ up in their ranking, then $x$ should still be a winner given $\mathbf{P}'$. Second, there is the variable-electorate axiom of \textit{positive involvement} \cite{Saari1995,Perez2001}: if a candidate $x$ is a winner given $\mathbf{P}$, and $\mathbf{P}^*$ is obtained from $\mathbf{P}$ by adding a new voter who ranks $x$ in first place, then $x$ should still be a winner given $\mathbf{P}^*$. These axioms are logically independent. For example, Instant Runoff Voting (see Section \ref{Methods}) satisfies positive involvement but not monotonicity; and a number of well-known voting methods satisfy monotonicity but not positive involvement.

There are at least two reactions to a voting method's violating monotonicity or positive involvement. One is that the method's violating one of these axioms is a serious problem if and only if violations are sufficiently frequent (according to some probability model). Another reaction is that the method's violating one of these axioms is a sign that the principle by which the voting method selects winners is fundamentally misconceived, which casts suspicion on its selection of winners in general, not just in those cases where the axiom is violated; but of course it is even worse for the method if in addition to having a misconceived principle for selecting winners, it frequently witnesses violations of the axiom.

In this paper, we discuss different ways of measuring the extent to which a voting method violates positive involvement (in particular, the eight different ways shown in Figure \ref{WaysTable}), using computer simulations. Our main aim is to determine appropriate measures of positive involvement violation and identify the main parameters affecting these measures---e.g., whether they are affected by variants of voting methods, numbers of candidates and voters, probability models, etc.---which is a necessary precursor to using positive involvement as part of an argument in favor of some voting methods over others in future work. 

First, we make some stage-setting conceptual points in Sections \ref{StrategySection}-\ref{ParticipationSection}, discuss related work in Section \ref{RelatedWork}, and recall technical preliminaries in Section \ref{PrelimSection}. Section \ref{QuantSection} contains our main discussion, as well as results of our simulations. Our methodological points in Section \ref{QuantSection} can be applied to measuring the extent of violation of other axioms in addition to positive involvement. We conclude in Section~\ref{Conclusion}.

\subsection{Strategic considerations}\label{StrategySection}

It is important to distinguish the perspective on positive involvement adopted above---we view it as an axiom, like monotonicity, that rules out perverse responses to unequivocal increases in support for a candidate---from a \textit{strategic perspective} on positive involvement, which is also adopted in the literature. A voting method's violating positive involvement may give a voter an incentive for \textit{strategic abstention}: if a voter $i$ knows that by casting a sincere ballot with her favorite candidate $a$ ranked first, this would kick $a$ out of the set of winners, then  $i$ may prefer not to vote rather than to cast a sincere ballot. This is so whenever $a$ would be the unique winner were $i$ not to vote but $a$ would not be were $i$ to vote sincerely. 

However, there are two important qualifications. First, if we are considering strategic abstention, we may also consider strategic \textit{voting}. For certain voting methods, whenever a candidate $a$ would be the unique winner were $i$ not to vote, there is always \textit{some} linear order with $a$ ranked first that $i$ can cast as her ballot to keep $a$ the unique winner,\footnote{When a voter casts a ballot with her true favorite $a$ ranked first but with deviations from her sincere preference lower down on the ballot in order to get $a$ elected, Dowding and Van Hees \cite{Dowding2008} call this a \textit{sincere manipulation}.} so $i$ has no incentive to abstain if she can vote strategically.\footnote{This is obvious if $i$ can submit as her ballot a strict weak order instead of a linear order, since then she may submit a fully indifferent ballot, which for many voting methods is equivalent to abstention, or a ballot with $a$ on top followed by an indifference class containing all other candidates. However, here we assume ballots are linear orders.} Second, we must consider the case where $a$ would not be the \textit{unique} winner were $i$ not to vote.  E.g., suppose that if $i$ were not to vote, the result would be a tie between $a$ and $c$, where $c$ is $i$'s least favorite candidate, whereas if $i$ were to vote sincerely, the unique winner would be $b$, who is $i$'s second favorite candidate.  Depending on $i$'s utility function, $i$ may well prefer $b$ to a tiebreaking process applied to $\{a,c\}$, in which case $i$ would prefer to vote sincerely rather than to abstain. So not every situation witnessing a failure of positive involvement is one in which the voter would prefer to abstain rather than vote sincerely.

Thus, the problem with violating positive involvement is not just that it may incentivize strategic abstention. Even in a society in which all voters always vote and always vote sincerely, so there is no risk of strategic abstention instead of sincere voting,  we still find violations of positive involvement perverse responses to unequivocal increases in support for a candidate. 

\subsection{Positive involvement vs. participation}\label{ParticipationSection}

It is also important to distinguish positive involvement from the axiom of \textit{participation} \cite{Moulin1988}. Participation is usually stated for \textit{resolute} voting methods that map each profile of voter preferences to a \textit{unique} winning candidate.\footnote{Resolute voting methods (defined on any profile of voter preferences, as in Definition \ref{VotingMethod}) either fail to treat voters equally---by violating the axiom of anonymity---or fail to treat candidates equally---by violating the axiom of neutrality (see \cite[\S~2.3]{Zwicker2016}).} A resolute voting method satisfies participation (and an arbitrary voting method satisfies what could be called \textit{resolute participation}) if adding a new voter who ranks $x$ above $y$ cannot result in a change from $x$ being the unique winner to $y$ being the unique winner.\footnote{For other definitions of participation for irresolute voting methods, see \cite{Perez2001,Jimeno2009,Sanver2012}.} Crucially, it is not required that the new voter ranks $x$ in first place. Thus, the new voter may rank other candidates above $x$, thereby hurting $x$'s prospects. Participation says that ranking those other candidates above $x$ should not  hurt $x$  more than it hurts $y$, so it should not result in $y$ becoming the unique winner. But this is not so clear in profiles containing majority cycles. In the presence of cycles, the main threat to $x$ may be a candidate $z$ who does not threaten $y$. Then a new voter with a ranking of the form $z\dots xy\dots$ may do more harm to $x$ than to $y$. This is only a sketch of an argument raising doubts about the participation axiom, but the key point is this: unlike violations of positive involvement, some violations of participation are not perverse responses to unequivocal increases in support for a candidate $x$, as some violations involve adding a voter who ranks other candidates above $x$. Of course, violations of participation give a voter an incentive to abstain rather than to vote sincerely, but that is a separate strategic issue that we set aside.

The term ``No Show Paradox'' was introduced by Fishburn and Brams \cite{Fishburn1983} for violations of what is now called \textit{negative involvement}.\footnote{The axiom of negative involvement \cite{Saari1995,Perez2001} states that adding a new voter who ranks a candidate last should not result in the candidate going from being a loser to a winner.  The analysis of this paper can also be applied to negative involvement, but for the sake of space we focus on positive involvement. } Later Moulin \cite{Moulin1988} changed the meaning of ``No Show Paradox'' to refer to violations of participation. Violations of positive and negative involvement are called instances of the ``Strong No Show Paradox''  by P\'{e}rez \cite{Perez2001}.\footnote{Plassman and Tideman \cite{Plassmann2014} use ``Strong No Show Paradox'' to refer to violations of participation, contradicting P\'{e}rez's use. P\'{e}rez \cite{Perez2001} calls violations of positive involvement the ``positive strong no show paradox'' and violations of negative involvement the ``negative strong no show paradox.'' Felsenthal and Tideman \cite{Felsenthal2013} and Felsenthal and Nurmi \cite{Felsenthal2016,Felsenthal2017} call them the ``P-TOP'' and ``P-BOT'' paradoxes, respectively.} P\'{e}rez  concludes that the Strong No Show Paradox is a common flaw of many \textit{Condorcet consistent} voting methods, which are methods  that always pick a Condorcet winner---a candidate who is majority preferred to every other candidate---if one exists. 

\subsection{Related Work}\label{RelatedWork}

This study fits into a line of recent work using computer simulations to estimate the frequency of violations of various voting criteria for different voting methods  \cite{Plassmann2014,Gehrlein2017,Brandt2019,Brandt2020}. The most closely related previous study is that of Brandt et al.~\cite{Brandt2019} on the frequency of the ``No Show Paradox'' in Moulin's sense \cite{Moulin1988}, i.e., the frequency of violations of the participation axiom. As the authors observe, ``While it is known that certain voting rules suffer from this paradox in principle, the extent to which it is of practical concern is not well understood'' (p.~520), a comment that also applies to violations of positive involvement. To fill the gap in the case of the No Show Paradox, Brandt et al.~use computer simulations---as well as analytic results obtained using Ehrhart theory---for six Condorcet consistent voting methods, three of which overlap with our list: Baldwin, Copeland, and Nanson. They find that ``for few alternatives, the probability of the NSP is rather small (less than 4\% for four alternatives and all considered preference models, except for Copeland's rule). As the number of alternatives increases, the NSP becomes much more likely and which rule is most susceptible to abstention strongly depends on the underlying distribution of preferences'' (p.~520). This observation about alternatives matches what we find for positive involvement, where the frequency of violations significantly increases with the number of alternatives.

\section{Preliminaries}\label{PrelimSection}

\subsection{Profiles}\label{Profiles}

Fix infinite sets $\mathcal{V}$ and $\mathcal{X}$ of \textit{voters} and \textit{candidates}, respectively. A given election will use only finite subsets $V\subseteq\mathcal{V}$ and $X\subseteq\mathcal{X}$. We consider elections in which each voter  submits a ranking of all the candidates, which we assume is a strict linear order. For a set $X$, let $\mathcal{L}(X)$ be the set of all strict linear orders on~$X$. For $P\in\mathcal{L}(X)$, we write `$xPy$' for $(x,y)\in P$; for $a\in X$, let $Rank(a,P)=|\{b\in X\mid bPa\}|+1$.

 \begin{definition}\label{ProfileDef}
 A \textit{profile} is a function $\mathbf{P}: V\to \mathcal{L}(X)$ for some nonempty finite $V\subseteq \mathcal{V}$, which we denote by $V(\mathbf{P})$ (called the set of \textit{voters in $\mathbf{P}$}), and nonempty finite $X\subseteq \mathcal{X}$,  which we denote by $X(\mathbf{P})$ (called the set of \textit{candidates in $\mathbf{P}$}). We call $\mathbf{P}(i)$ voter $i$'s \textit{ballot} and write `$\mathbf{P}_i$' for $\mathbf{P}(i)$.
\end{definition}

\begin{definition}\label{DisjointUnion} For a profile $\mathbf{P}$ and set $C\subseteq V(\mathbf{P})$ of voters, let $\mathbf{P}_{-C}$ be the restriction of $\mathbf{P}$ to $V(\mathbf{P})\setminus C$. For  a single voter $i\in V(\mathbf{P})$, let $\mathbf{P}_{-i}=\mathbf{P}_{-\{i\}}$.  Given profiles $\mathbf{P}$ and $\mathbf{P}'$ such that $X(\mathbf{P})=X(\mathbf{P}')$ and $V(\mathbf{P})\cap V(\mathbf{P}')=\varnothing$, we define the profile $\mathbf{P}+\mathbf{P}': V(\mathbf{P})\cup V(\mathbf{P}')\to \mathcal{L}(X(\mathbf{P}))$ as follows: if $i\in V(\mathbf{P})$, then  $(\mathbf{P}+\mathbf{P}')(i) = \mathbf{P}(i)$, and if $i\in V(\mathbf{P}')$, then     $(\mathbf{P}+\mathbf{P}')(i) = \mathbf{P}'(i)$. 
\end{definition}

\subsection{Voting Methods}\label{Methods}

\begin{definition}\label{VotingMethod} A \textit{voting method} is a function $F$ on the domain of all profiles such that for any profile $\mathbf{P}$, ${\varnothing\neq F(\mathbf{P})\subseteq X(\mathbf{P})}$.
\end{definition}
In this paper, we consider the following voting methods. We chose these methods because they give a broad representation of different classes of methods (e.g., scoring rules, iterative methods, and Condorcet methods, including C1 and C2 methods) and can be efficiently computed (except for Ranked Pairs).

 \textbf{Positional scoring rules.}  A \textit{scoring vector} is a vector  $\langle s_1, \ldots, s_n\rangle$ of numbers such that for each ${m\in \{1,\ldots, n-1\}}$, $s_m \ge s_{m+1}$. Given a profile $\mathbf{P}$ with $|X(\mathbf{P})|=n$, $x\in X(\mathbf{P})$, a scoring vector $\vec{s}$ of length $n$, and $i\in V(\mathbf{P})$, define $score_{\vec{s}}(x, \mathbf{P}_i)=s_r$ where $r=Rank(x,\mathbf{P}_i)$. Let $score_{\vec{s}}(x,\mathbf{P})= \sum_{i\in V(\mathbf{P})} score_{\vec{s}}(x,\mathbf{P}_i)$. A voting method $F$ is a \textit{positional scoring rule} if there is a map $\mathcal{S}$ assigning to each natural number $n$ a scoring vector of length $n$ such that for any profile $\mathbf{P}$ with $|X(\mathbf{P})|=n$, $F(\mathbf{P})=\mathrm{argmax}_{x\in X(\mathbf{P})} score_{\mathcal{S}(n)}(x,\mathbf{P})$.  Two well-known examples of scoring rules are:
\begin{center}
\textbf{Plurality}: $\mathcal{S}(n)=\langle 1,0,\dots,0\rangle$ \quad and \quad \textbf{Borda}: $\mathcal{S}(n)=\langle n-1,n-2,\dots, 1,0\rangle$.
\end{center}

 $\textbf{Instant Runoff}$ (also known as Alternative Vote, Ranked Choice, and Single Transferable Vote): Iteratively remove all candidates with the fewest number of voters who rank them first, until there is a candidate who is a majority  winner (i.e., ranked first by a strict majority of voters).\footnote{When there is more than one candidate with the fewest number of voters who rank them first, this definition of Instant Runoff, taken from \cite[p.~7]{Taylor2008}, eliminates all of them. For another way of handling such ties, see \cite{Wangetal2019} and Footnote \ref{IRPUT}.} If, at some stage of the removal process, all remaining candidates have the same number of  voters who rank them first (so all candidates would be removed), then all remaining candidates  are selected as winners.

 $\textbf{Coombs}$ \cite{Coombs1964,Grofman2004}:  Iteratively remove all candidates with the most  number of voters who rank them last, until there is a candidate who is a majority  winner.   If, at some stage of the removal process, all remaining candidates have the same number voters who rank them last (so all candidates would be removed), then all remaining candidates  are selected as winners.

 $\textbf{Baldwin}$  \cite{Baldwin1926}:  Iteratively remove all candidates with the smallest Borda score, until there is a single candidate remaining.   If, at some stage of the removal process, all remaining candidates have the same Borda score (so all candidates would be removed), then all remaining candidates  are selected as winners.

$\textbf{Nanson}$ \cite{Nanson1882}: There are two versions of this voting method \cite{Niou1987}. \textbf{Strict Nanson} (resp.~\textbf{Weak Nanson}) iteratively removes all candidates whose Borda score is strictly less than (resp.~less than or equal to) the average Borda score of the candidates remaining at that stage, until one candidate remains. If, at some stage of the removal process, all remaining candidates have the same Borda score (so all candidates would be removed), then all remaining candidates  are selected as winners. Although Nanson seems to have intended \textbf{Weak Nanson} (see \cite{Niou1987}), the literature on computational social choice usually interprets `Nanson' as \textbf{Strict Nanson} (see, e.g., \cite{Narodytska2011,Zwicker2016,Fischer2016,Conitzer2016}), so we focus on \textbf{Strict Nanson} in this paper.

\textbf{Bucklin} \cite{Hoag1926}: Given a candidate $a$ in a profile $\mathbf{P}$ and a positive integer $n$, say that $a$ is an \textit{$n$-th level majority winner} in $\mathbf{P}$ if a strict majority of voters rank $a$ in $n$-th place or higher (thus, a majority winner in the usual sense is a 1st level majority winner). Where $k$ is the smallest positive integer for which there is at least one $k$-th level majority winner, Bucklin selects as winners the $k$-th level majority winners for whom the most voters rank them in $k$-th place or higher.\footnote{\label{SimplifiedBucklin}The \textbf{Simplified Bucklin} method selects as winners all $k$-th level majority winners. We ran our simulations with Simplified Bucklin as well as Bucklin; its frequency of violating positive involvement is similar to Bucklin but slightly worse.} 

The other methods that we study in this paper are from a broad class of so-called majority or margin-based methods.   We need the following notation for defining these methods.

\begin{definition} For a profile $\mathbf{P}$ and $a,b\in X(\mathbf{P})$, let~$Margin_\mathbf{P}(a,b)=|\{i\in V(\mathbf{P})\mid a\mathbf{P}_ib\}| -|\{i\in V(\mathbf{P})\mid b\mathbf{P}_i a\}|$ be the margin of $a$ over $b$ in $\mathbf{P}$.   The \textit{margin graph of $\mathbf{P}$} is the weighted directed graph whose set of vertices is $X(\mathbf{P})$ with an edge from $a$ to $b$ when $Margin_\mathbf{P}(a,b)>0$, weighted by $Margin_\mathbf{P}(a,b)$.
\end{definition}

\textbf{Copeland} \cite{Copeland1951}: The Copeland score of $a\in X(\mathbf{P})$ is the number of $b\in X(\mathbf{P})$ such that $Margin_\mathbf{P}(a,b)>0$ minus the number of $b\in X(\mathbf{P})$ such that $Margin_\mathbf{P}(b,a)>0$. Then $Copeland(\mathbf{P})$ is the set of  $x\in X(\mathbf{P})$ with maximal Copeland score.\footnote{\label{LlullNote}An equivalent definition of Copeland defines the score of $a\in X(\mathbf{P})$ as the number of ${b\in X(\mathbf{P})}$ with $Margin_\mathbf{P}(a,b)>0$ plus $1/2$ times the number of ${b\in X(\mathbf{P})}$ with $Margin_\mathbf{P}(a,b)=0$. We also ran all of our simulations using the variant of Copeland known as Llull, where the score of $a\in X(\mathbf{P})$ is the number of ${b\in X(\mathbf{P})}$ with $Margin_\mathbf{P}(a,b)>0$ plus the number of ${b\in X(\mathbf{P})}$ with $Margin_\mathbf{P}(a,b)=0$ (see \cite{Faliszewski2007}). Llull's performance was very similar to that of Copeland.}

 \textbf{Top Cycle} \cite{Smith1973,Schwartz1986} (also known as Smith and GETCHA):  For a profile $\mathbf{P}$, let  $a \succsim_\mathbf{P} b$ if $Margin_\mathbf{P}(a,b)\geq 0$. Let $\succsim_\mathbf{P}^*$ be the transitive closure of $\succsim_\mathbf{P}$.  Then $Top Cycle(\mathbf{P})=\{x\in X(\mathbf{P})\mid \mbox{for all }y\in X(\mathbf{P}), x\succsim_\mathbf{P}^*y\}$.\footnote{We also ran our simulations on the GOCHA method \cite{Schwartz1986} that differs from Top Cycle only in profiles with margins of 0 between some candidates. Its performance was similar or slightly better than that of Top Cycle, depending on the parameters of the simulation.}

 \textbf{Uncovered Set} (Gillies version) \cite{Gillies1959}: Given $a,b\in X(\mathbf{P})$,   $a$ \textit{Gillies covers} $b$ in $\mathbf{P}$ if $Margin_\mathbf{P}(a,b)>0$ and for all $c\in X(\mathbf{P})$, if   $Margin_\mathbf{P}(c,a)>0$, then $Margin_\mathbf{P}(c,b)>0$. Then $UC(\mathbf{P})$ is the set of candidates  who are not Gillies covered in $\mathbf{P}$. This is called the Gillies Uncovered Set in \cite{Duggan2013}.
 
 \textbf{Ranked Pairs} \cite{Tideman1987}: For a profile $\mathbf{P}$ and $T\in \mathcal{L}\big(\{(x,y)\mid x\neq y\mbox{ and }Margin_\mathbf{P}(x,y)\geq 0 \}\big)$, called the \textit{tie-breaking ordering}, a pair $(x,y)$ of candidates has a \textit{higher priority} than a pair $(x',y')$ of candidates according to $T$ when either  $Margin_\mathbf{P}(x,y) > Margin_\mathbf{P}(x',y')$ or $Margin_\mathbf{P}(x,y) = Margin_\mathbf{P}(x',y')$ and $(x,y)\mathrel{T} (x',y')$.  We construct a \textit{Ranked Pairs ranking} $\succ_{\mathbf{P},T}\,\in \mathcal{L}(X)$ as follows:
\begin{enumerate}
\item Initialize $\succ_{\mathbf{P},T}$ to $\varnothing$.
\item If all pairs $(x,y)$ with $x\neq y$ and $Margin_\mathbf{P}(x,y)\geq 0$ have been considered, then return $\succ_{\mathbf{P},T}$.  Otherwise let $(a,b)$ be the pair with the highest priority  among those with $a\neq b$ and $Margin_\mathbf{P}(a,b)\geq 0$ that have not been considered so far.  
\item If $\succ_{\mathbf{P},T}\cup\, \{(a,b)\}$ is acyclic, then add $(a,b)$ to $\succ_{\mathbf{P},T}$; otherwise, add $(b,a)$ to $\succ_{\mathbf{P},T}$.   Go to step 2. 
\end{enumerate}
When the procedure terminates, $\succ_{\mathbf{P},T}$ is a linear order. The set $RP(\mathbf{P})$ of Ranked Pairs winners is the set of all $x\in X(\mathbf{P})$ such that $x$ is the maximum of $\succ_{\mathbf{P},T}$ for some tie-breaking ordering $T$. This is the ``parallel universe'' version of Ranked Pairs called \textsf{RP} in \cite{BrillFischer2012,Wangetal2019}, distinguished from other non-neutral, non-anonymous, or probabilistic versions of Ranked Pairs. 

Since calculating $RP(\mathbf{P})$ is an NP-complete problem \cite{BrillFischer2012}, we also consider the non-anonymous version of Ranked Pairs proposed by Zavist and Tideman \cite{ZavistTideman1989}, which we call  \textbf{Ranked Pairs ZT}. Zavist and Tideman propose to use a distinguish voter's ranking to derive the tie-breaking ordering $T$. In particular, given $i\in V(\mathbf{P})$, let $T(\mathbf{P}_i)$ be the lexicographic order on $\{(x,y)\mid x\neq y\mbox{ and }Margin_\mathbf{P}(x,y)\geq 0 \}$ derived from $\mathbf{P}_i$. Since different profiles have different sets of voters, we cannot use the same distinguished voter for all profiles. Given a linear order $L$ of $\mathcal{V}$, for any profile $\mathbf{P}$, we define $RPZT_L(\mathbf{P})$ to be the set of all $x\in X(\mathbf{P})$ such that $x$ is the maximum of $\succ_{\mathbf{P},T(\mathbf{P}_i)}$ where $i$ is the minimal element of $V(\mathbf{P})$ according to $L$.
 
  \textbf{Beat Path} \cite{Schulze2011}: For $a,b\in X(\mathbf{P})$, a \textit{path from $a$ to $b$ in $\mathbf{P}$} is a sequence $\rho=x_1,\dots,x_n$ of distinct candidates in $X(\mathbf{P})$ with $x_1=a$ and $x_n=b$ such that for $1\leq k\leq n-1$, $Margin_\mathbf{P}(x_k,x_{k+1})>0$. The \textit{strength of $\rho$} is $\mbox{min}\{Margin_\mathbf{P}(x_k,x_{k+1})\mid 1\leq k\leq n-1\}$. Then  $a$ defeats $b$ in $\mathbf{P}$ according to Beat Path if the strength of the strongest path from $a$ to $b$ is greater than the strength of the strongest path from $b$ to~$a$.  $BP(\mathbf{P})$ is the set of undefeated candidates. 

 \textbf{Split Cycle} \cite{HP2020}: A \textit{majority cycle in $\mathbf{P}$} is a sequence $\rho=x_1,\dots,x_n$ of distinct candidates in $X(\mathbf{P})$ except $x_1=x_n$ such that for $1\leq k\leq n-1$, $Margin_\mathbf{P}(x_k,x_{k+1})>0$. The \textit{strength of} $\rho$ is defined as above for Beat Path. Then  $a$ defeats $b$ in $\mathbf{P}$ according to Split Cycle if $Margin_\mathbf{P}(a,b)$ is positive and greater than the strength of the strongest majority cycle containing $a$ and $b$. $SC(\mathbf{P})$ is the set of undefeated candidates.

\subsection{Positive Involvement}

As explained in Section \ref{Introduction}, our interest in this paper is the axiom of positive involvement.

\begin{definition}\label{InvolveDef} A voting method $F$ satisfies \textit{positive involvement} (PI) if for any profile $\mathbf{P}$ and ${x\in F(\mathbf{P})}$, if $\mathbf{P}'$ is obtained from $\mathbf{P}$ by adding a new voter who ranks $x$ in first place, then $x\in F(\mathbf{P}')$.
\end{definition}

\noindent PI is sometimes discussed in terms of the addition of several voters who rank $x$ first. An obvious inductive argument shows that the coalitional version of PI follows from the single-voter version in Definition \ref{InvolveDef}.

\begin{lemma} If a voting method $F$ satisfies PI, then it satisfies \emph{coalitional} PI: for any profile $\mathbf{P}$ and $x\in F(\mathbf{P})$, if $\mathbf{P}'$ is a profile with $X(\mathbf{P})=X(\mathbf{P}')$, $V(\mathbf{P})\cap V(\mathbf{P}')=\varnothing$, and every voter in $\mathbf{P}'$ ranks $x$ in first place, then $x\in F(\mathbf{P}+\mathbf{P}')$.
\end{lemma}

The voting methods in Section \ref{Methods} can be classified by their satisfying or violating PI as follows.

\begin{proposition}\label{MethodsViolatePI} The voting methods  Baldwin, Beat Path, Bucklin, Coombs, Copeland, Ranked Pairs, Strict Nanson, Top Cycle, and Uncovered Set all violate PI---see the Appendix for examples.  All positional scoring rules, Instant Runoff, and Split Cycle satisfy PI (see \cite{Perez2001,HP2020}).\footnote{\label{RunoffNote}Another well-known method satisfying PI is Plurality with Runoff \cite[\S~4.7.1]{Felsenthal2017}. This result depends on a specific way of breaking ties when more than two candidates qualify for the runoff (a candidate $a$ qualifies for the runoff if either $a$ has maximal plurality score or there is a unique candidate with maximal plurality score and $a$ is among the candidates with the second highest plurality score). The version of Plurality with Runoff satisfying PI is the ``parallel universe'' version that considers all possible duels between candidates who qualify for the runoff; a candidate is a winner just in case they win in one of these duels. If instead all candidates who qualify for the runoff are promoted to a second round decided by Plurality, then the resulting version of Plurality with Runoff does not satisfy PI. Thanks to an anonymous referee for discussion of these points.} 
\end{proposition}

\begin{remark} For results on violation or satisfaction of PI by other voting methods, see \cite{Perez2001,Felsenthal2017}. Note that few known voting methods satisfy both PI and Condorcet consistency. Indeed, P\'{e}rez  \cite{Perez2001} observed that with the exception of the Minimax method \cite{Simpson1969,Kramer1977}, which satisfies PI, ``all the Condorcet correspondences that (to the best of our knowledge) are proposed in the literature'' violate PI (p.~601). Other examples of Condorcet methods violating PI, besides those listed in Proposition \ref{MethodsViolatePI} (Baldwin, Beat Path, Copeland, Ranked Pairs, Strict Nanson, Top Cycle, and Uncovered Set), are Dodgson \cite{Dodgson1876}, Kemeny \cite{Kemeny1959}, and Young \cite{Young1977} (see \cite{Perez2001,Felsenthal2017}).\footnote{We did not include Dodgson, Kemeny, and Young in our simulations due to the computational complexity of determining winners for these methods (see \cite{Caragiannis2016} and \cite[\S~4.2]{Fischer2016}).} However, Split Cycle is Condorcet consistent and satisfies PI.\end{remark}

\section{Quantitative Analysis}\label{QuantSection}

We now turn to our discussion of different ways to quantify the extent to which a voting method $F$ violates PI. They are summarized in Figure \ref{WaysTable} and explained in the subsections to follow. The most obvious idea is to simply consider the probability that a randomly drawn profile $\mathbf{P}$ (according to some probability model) \textit{witnesses a violation of PI for $F$}, meaning that there is some voter $i$ in $\mathbf{P}$ such that $i$'s favorite candidate wins in $\mathbf{P}_{-i}$ according to $F$ but not in $\mathbf{P}$. However, for medium to large sized electorates, we should expect this probability to be low for the mundane reason that the addition of any \textit{single} voter will rarely cause someone to lose. We will explore several ways to deal with this issue. 

One natural idea is to look at violations of \textit{coalitional} PI: say that a profile $\mathbf{P}$ \textit{witnesses a violation of coalitional PI for  $F$} if there is some coalition $C$ of voters in $\mathbf{P}$ with the same ranking (or at least the same top-ranked candidate), and their favorite candidate wins in $\mathbf{P}_{-C}$ but not in $\mathbf{P}$ according to $F$. Ideally, we would like to estimate the probability that a random profile witnesses a violation of coalitional PI (for various coalition sizes). Unfortunately, we find it too computationally expensive to check for every coalition of voters whether removing that coalition shows a violation of coalitional PI, especially for the thousands of profiles needed for reliable simulation results. To deal with this problem, in Section \ref{profile-coalitionparadigm} we will assess the probability that a random \textit{profile-coalition pair} witnesses a violation of PI.

However, we will begin in Section \ref{profileparadigm} with the probability that a random profile witnesses a violation of single-voter PI, since this provides a baseline with which to compare everything else that follows.

The probability model for profiles that will serve as our baseline is the Impartial Culture (IC) model: when sampling profiles with $n$ candidates and $m$ voters, the IC model gives every such profile equal probability. In Section \ref{OtherProbModels}, we  consider several other probability models for sampling profiles. For each data point in our graphs, we sampled 25,000 profiles with the indicated even number $m$ of voters and 25,000 profiles with the indicated odd number $m+1$ of voters, in order to have a mix of even and odd-sized electorates. We used the preflib \cite{Mattei2013} implementation for each of the probability models. Our code is in the online supplementary material at \url{https://github.com/epacuit/posinvolvement}.
\begin{figure}
\begin{center}
\begin{tabular}{c|c|c}
& profiles & profile-coalition pairs \\
\hline 
absolute frequency & column 1 of Figure \ref{Absolute+ConditionalRemoval} & column 1 of Figure \ref{Absolute+ConditionalAddition} \\
conditional on disagreement with $F_2$ & columns 2-4 of Figure \ref{Absolute+ConditionalRemoval} & columns 2-4 of Figure \ref{Absolute+ConditionalAddition}  \\
relativized to voter potency & column 1 of Figure \ref{Absolute+ConditionalRemovalRelativized} & column 1 of Figure \ref{Absolute+ConditionalRelativizedAddition} \\
conditional on disagreement \& relativized & columns 2-4 of Figure \ref{Absolute+ConditionalRemovalRelativized} & columns 2-4 of Figure \ref{Absolute+ConditionalRelativizedAddition} 
\end{tabular}
\end{center}
\caption{ways of measuring PI violation}\label{WaysTable}
\end{figure}

\subsection{Random profiles}\label{profileparadigm}

\subsubsection{Probability of PI violation}

The leftmost column in Figure \ref{Absolute+ConditionalRemoval} shows, for several voting methods $F$ that violate PI, the estimated probability that a random profile (according to IC) witnesses a violation of PI for $F$, as defined above. The important qualitative observations are that (i) the probability of PI violation increases as the number of candidates increases and (ii) the probability of PI violation decreases as the number of voters increases (above 20), for the mundane reason mentioned above---as the number of voters increases, any single voter has less influence in the election (a point to which we return in Section \ref{ProfileRelativize}).

\subsubsection{Probability of PI violation conditional on voting method disagreement}\label{ProfileCondition}

\begin{figure*}[!htb]
\centering
\includegraphics[scale=0.416]{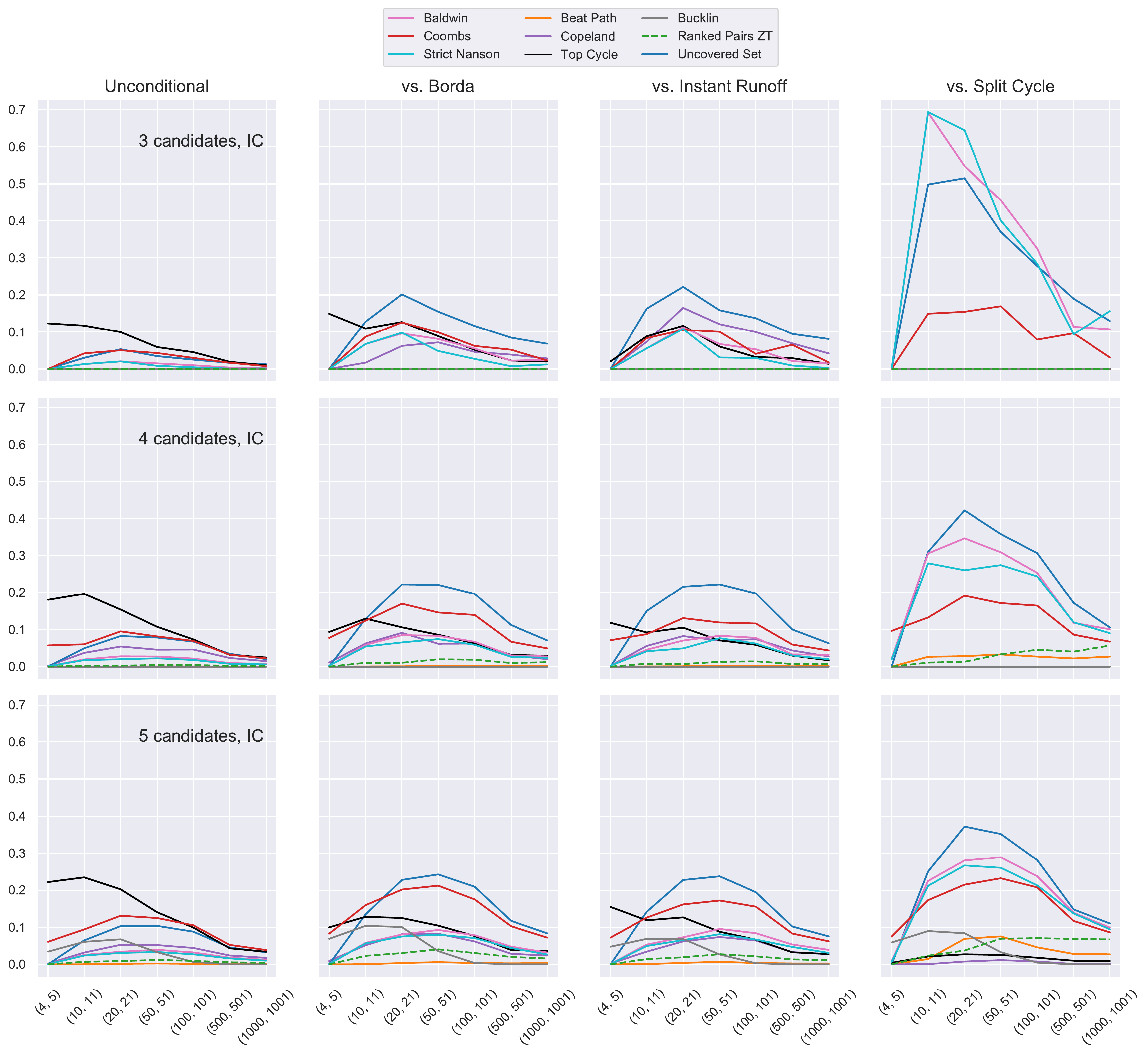}

\caption{probability of a profile $\mathbf{P}$ witnessing a violation of PI for $F$ either unconditionally (far left), conditional on $F$ disagreeing with Borda in $\mathbf{P}$ (second from left), conditional on $F$ disagreeing with Instant Runoff  in $\mathbf{P}$ (second from right), or conditional on $F$ disagreeing with Split Cycle in $\mathbf{P}$  (far right). Ranked Pairs was not included due to computational limitations (cf.~\cite{BrillFischer2012}). Ranked Pairs ZT is shown with a dashed line to mark it out as our only non-anonymous voting method.}\label{Absolute+ConditionalRemoval}
\end{figure*}

Although the absolute frequency of PI violation in the leftmost column of Figure \ref{Absolute+ConditionalRemoval} is relevant, it is a mistake to think that if a voting method $F_1$ has a low frequency of violating PI, then this undermines the use of PI as a reason to favor another voting method $F_2$, which satisfies PI, over $F_1$. First of all, when using PI to help choose between a method $F_1$ that violates the axiom and a method $F_2$ that satisfies it, we should ask: \textit{in the profiles in which $F_1$ and $F_2$ disagree}, with what frequency does $F_1$ violate PI?\footnote{We think that this methodology (which, as far as we know, is new) should be applied to the study of voting axioms in general, not only to PI.} That is, we should consider the following conditional probability, for a random profile $\mathbf{P}$ (according to the given probability model, for a given number of candidates and voters):
\begin{equation}
Pr(\mathbf{P}\mbox{ witnesses a violation of PI for }F_1 \mid F_1(\mathbf{P})\neq F_2(\mathbf{P})).\label{ConditionEq}
\end{equation}
Columns 2-4 of Figure \ref{Absolute+ConditionalRemoval} show estimates of this conditional probability for several voting methods $F_1$ that violate PI compared to three choices for $F_2$: Borda (2nd column), Instant Runoff (3rd column), and Split Cycle (4th column). We see a striking increase in the probability that $F_1$ will violate PI when we condition on $F_1$ disagreeing with the PI-satisfying method $F_2$. Note that if the probability of $F_1$ violating PI is higher conditional on $F_1$ disagreeing with $F_2$ than it is conditional on $F_1$ disagreeing with $F_2'$, then PI provides a stronger argument in the context of deciding between $F_1$ and $F_2$ than it does in the context of deciding between $F_1$ and $F_2'$.  E.g., let $F_1$ be Baldwin, $F_2$ be Split Cycle, and $F_2'$ be Borda. 

\begin{remark}\label{TiesRemark} Since violation of (single voter) PI requires a close election in order for one voter to affect the outcome, we checked (i) to what extent the frequency of PI violation by Baldwin and Coombs depends on the manner of handling ties at intermediate stages of their elimination procedures and (ii) to what extent the frequency of PI violation by Uncovered Set depends on which of several variants of the Uncovered Set---which can disagree only when some candidates have a margin of 0---one chooses to analyze. 

The tie-handling issue arises for Baldwin (resp.~Coombs) in case there are multiple candidates with the smallest Borda score (resp.~most last place votes) at a given stage of iteration. For each iterative elimination method $F$, there is a ``parallel-universe tie-handling'' (PUT) variant (cf.~\cite{Freeman2015}): a candidate $a$ wins under the PUT variant of $F$ just in case there is some linear order $L$ on the set of candidates such that $a$ wins according to $F_L$, where $F_L$ is defined in the same was as $F$ except that if multiple candidates meet the criterion for elimination at some stage (e.g., have the smallest Borda score, or the most last place votes), only the $L$-minimal candidate among those candidates is eliminated. We found that the PUT versions of Baldwin and Coombs performed similarly or slightly better than the versions defined in Section \ref{Methods}.\footnote{\label{IRPUT}We also checked (for 3 and 4 candidates) the difference between conditioning on disagreement with Instant Runoff, as in Figure \ref{Absolute+ConditionalRemoval}, and conditioning on disagreement with the PUT version of Instant Runoff. We found the results to be similar.} But given the computational difficulty of determining winners for the PUT versions, we use the versions defined in Section \ref{Methods} for the rest of our analysis.

As for Uncovered Set, we considered three variants in addition to the Gillies Uncovered Set in Section \ref{Methods}: the Bordes Uncovered Set, the McKelvey Uncovered Set, and the Fishburn Uncovered Set.\footnote{Following Duggan \cite{Duggan2013}, say that $a$ \textit{Bordes covers} $b$ in $\mathbf{P}$ if $Margin_\mathbf{P}(a,b)>0$ and for all $c\in X(\mathbf{P})$, if   $Margin_\mathbf{P}(c,a)\geq 0$, then $Margin_\mathbf{P}(c,b)\geq 0$; and say that $a$ \textit{McKelvey covers} $b$ if $a$ Gillies covers and Bordes covers $b$.  For Fishburn's \cite{Fishburn1977} variant, say that $a$ \textit{Fishburn covers} $b$ in $\mathbf{P}$ if for all $c\in X(\mathbf{P})$, if   $Margin_\mathbf{P}(c,a)> 0$, then $Margin_\mathbf{P}(c,b)> 0$, and there is a $c\in X(\mathbf{P})$ such that $Margin_\mathbf{P}(c,b)> 0$ and $Margin_\mathbf{P}(c,a)\leq 0$.} Although the absolute frequencies of PI violation for these variants are similar (with Gillies and Fishburn almost indistinguishable, and Bordes and McKelvey almost indistinguishable), the frequencies of PI violation conditional on disagreement with Borda, Instant Runoff, and Split Cycle are surprisingly different, with Gillies far worse than the others (see our online supplementary material). In this paper, we focus on the Gillies variant for a worst-case analysis of the family of Uncovered Set variants.\end{remark}

\subsubsection{Probability of PI violation relativized to voter potency}\label{ProfileRelativize}

Although the results of conditioning on voting method disagreement are striking, even those conditional probabilities are bound to decline as we increase the number of voters, given a single voter's declining influence. What we should be asking, we think, is how likely a PI violation is compared to how likely it is that any single voter causes a candidate to lose. For this, we introduce the notion of voter \textit{potency}.

\begin{definition}\label{PotencyDef} For any profile $\mathbf{P}$ and $i\in V(\mathbf{P})$, we say that \textit{$i$ is potent in $\mathbf{P}$} if $F(\mathbf{P}_{-i})\not\subseteq F(\mathbf{P})$.\footnote{Note that $i$ being potent in $\mathbf{P}$ is stronger than $i$ being \textit{pivotal} in $\mathbf{P}$ in the sense that $F(\mathbf{P}_{-i})\neq F(\mathbf{P})$.}
\end{definition}

\begin{figure*}[!htb]
\centering
\includegraphics[scale=0.445]{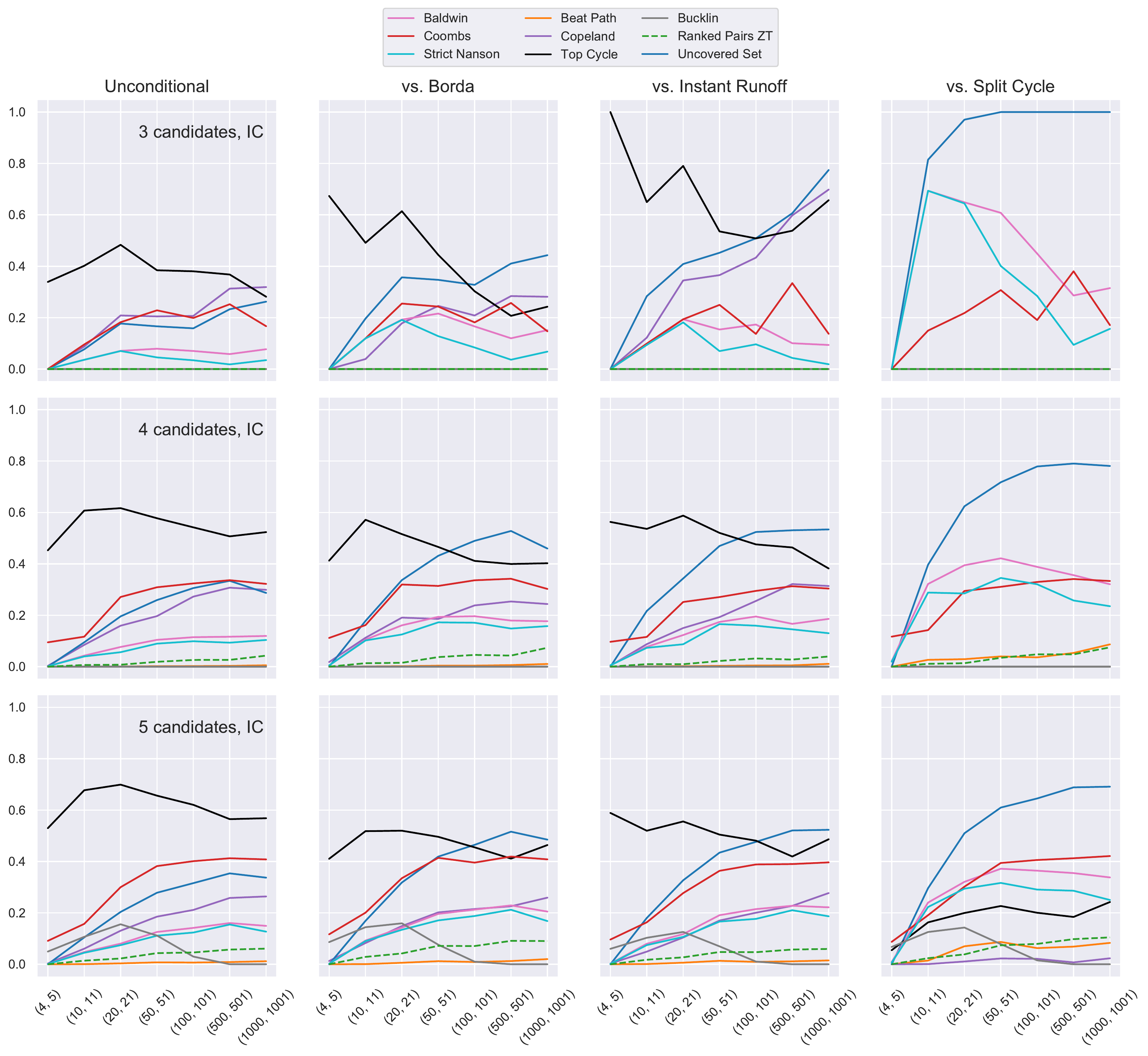}

\caption{probability of a profile $\mathbf{P}$ witnessing a violation of PI for $F$, divided by the probability of $\mathbf{P}$ having a potent voter, either unconditionally (far left), conditional on $F$ disagreeing with Borda in $\mathbf{P}$ (second from left), conditional on $F$ disagreeing with Instant Runoff in $\mathbf{P}$ (second from right), or conditional on $F$ disagreeing with Split Cycle in $\mathbf{P}$ (far right).}\label{Absolute+ConditionalRemovalRelativized}
\end{figure*}

Then our proposal is to measure how badly a voting method violates single-voter PI by the ratio
\begin{equation}\frac{Pr( \mbox{there is a voter }i\mbox{ triggering a PI violation, i.e., }\mathrm{max}(\mathbf{P}_i)\in F_1(\mathbf{P}_{-i})\setminus F_1(\mathbf{P}))}{Pr(\mbox{there is a voter }i\mbox{ who is potent in }\mathbf{P})},\end{equation}
possibly with these probabilities conditioned on $F_1$ disagreeing in $\mathbf{P}$ with an $F_2$ that satisfies PI. For voting methods violating PI, the numerator may be small (as opposed to 0 for methods that satisfy PI), but the denominator is also small; thus, the ratio may be surprisingly large. The results are shown in Figure \ref{Absolute+ConditionalRemovalRelativized}.   

\begin{remark}\label{ConditioningRemark}One curious feature of the last column of Figure \ref{Absolute+ConditionalRemovalRelativized} is that for 3 or 4 candidates, the probability of  Copeland and Top Cycle violating PI conditional on their disagreeing with Split Cycle in $\mathbf{P}$ goes \textit{down to zero}. This is initially puzzling: since Split Cycle satisfies PI, mustn't Copeland and Top Cycle disagree with Split Cycle whenever they violate PI? The answer is `yes', but they may disagree with Split Cycle in the \textit{smaller profile} $\mathbf{P}_{-i}$ rather than in $\mathbf{P}$; indeed, this always happens when the cited methods violate single-voter PI for 3 or 4 candidates. Thus, there is an ambiguity in the idea of ``conditioning on disagreement with $F_2$''---we could condition on disagreement in $\mathbf{P}_{-i}$, in $\mathbf{P}$, in both, or in one or the other. So far we have only conditioned on disagreement in $\mathbf{P}$, since so far in our random sampling we only draw a profile $\mathbf{P}$; but in Section \ref{ProfileCoalitionDisagree}, when we randomly sample profile-voter pairs, we will be able to conveniently condition on disagreement in $\mathbf{P}_{-i}$ or $\mathbf{P}$.\end{remark}

\subsection{Random profile-coalition pairs}\label{profile-coalitionparadigm}

Given the computational difficulty of checking whether a profile witnesses a violation of coalitional PI, we now turn to randomly sampling \textit{profile-coalition pairs}, i.e., pairs  $(\mathbf{P},\mathbf{P}_C)$ of a random profile $\mathbf{P}$ (for a given number of candidates and voters) and a ``coalitional'' profile $\mathbf{P}_C$ obtained by randomly selecting a single ranking and then assigning it to all voters in $C$ (with $C\cap V(\mathbf{P})=\varnothing$). Such a profile-coalition pair \textit{witnesses a violation of PI} just in case the unanimously top-ranked candidate in $\mathbf{P}_C$ is a winner in $\mathbf{P}$ but not in $\mathbf{P}+\mathbf{P}_C$. Note that even if the pair $(\mathbf{P},\mathbf{P}_C)$ does not witness a violation of PI, the profile $\mathbf{P}+\mathbf{P}_C$ may witness a violation of coalitional PI by the removal of a different coalition $C'$. Thus, this approach misses violations of coalitional PI involving other coalitions. The probability that a random profile-coalition pair witnesses a violation of PI may be considerably lower than the probability that a random profile witnesses a violation of coalitional PI. Nonetheless, there are benefits of this approach, such as (i) our being able to feasibly study coalitions of more than one voter, (ii) our being able to search profiles up to 5,000 voters or up to 10 candidates, and (iii) our being able to conveniently condition on voting method disagreement before or after the new coalition of voters joins the election (see Remark~\ref{ConditioningRemark}).

\subsubsection{Probability of PI violation}

The leftmost column of Figure \ref{Absolute+ConditionalAddition} shows estimates for the probability that a randomly selected profile-coalition pair $(\mathbf{P},\mathbf{P}_C)$ witnesses a violation of PI, depending on whether $C$ is a coalition of a single voter (first row), a coalition of 0.25\% of the total voter size (second row), a coalition of 0.5\% (third row), or a coalition of 0.75\% (fourth row). We show only the results for 500, 1,000, and 5,000 voter profiles, since with 100 voters or fewer, all the coalition sizes round to 1 voter. As expected, in the single voter case, the probability that a random profile-voter pair witnesses a violation of PI is much lower than the probability that a random profile witnesses a violation of single-voter PI (compare the leftmost column and third row of Figure \ref{Absolute+ConditionalRemoval}). Indeed, the difference is roughly an order of magnitude. The probability does increase for larger coalition sizes, but the probability of a random profile-0.5\%-coalition pair witnessing a violation of PI is still less than half the probability of a random profile $\mathbf{P}$ witnessing a violation of single-voter PI.

\subsubsection{Probability of PI violation conditional on disagreement}\label{ProfileCoalitionDisagree}

\begin{figure*}[!htbp]
\centering
\includegraphics[scale=0.4]{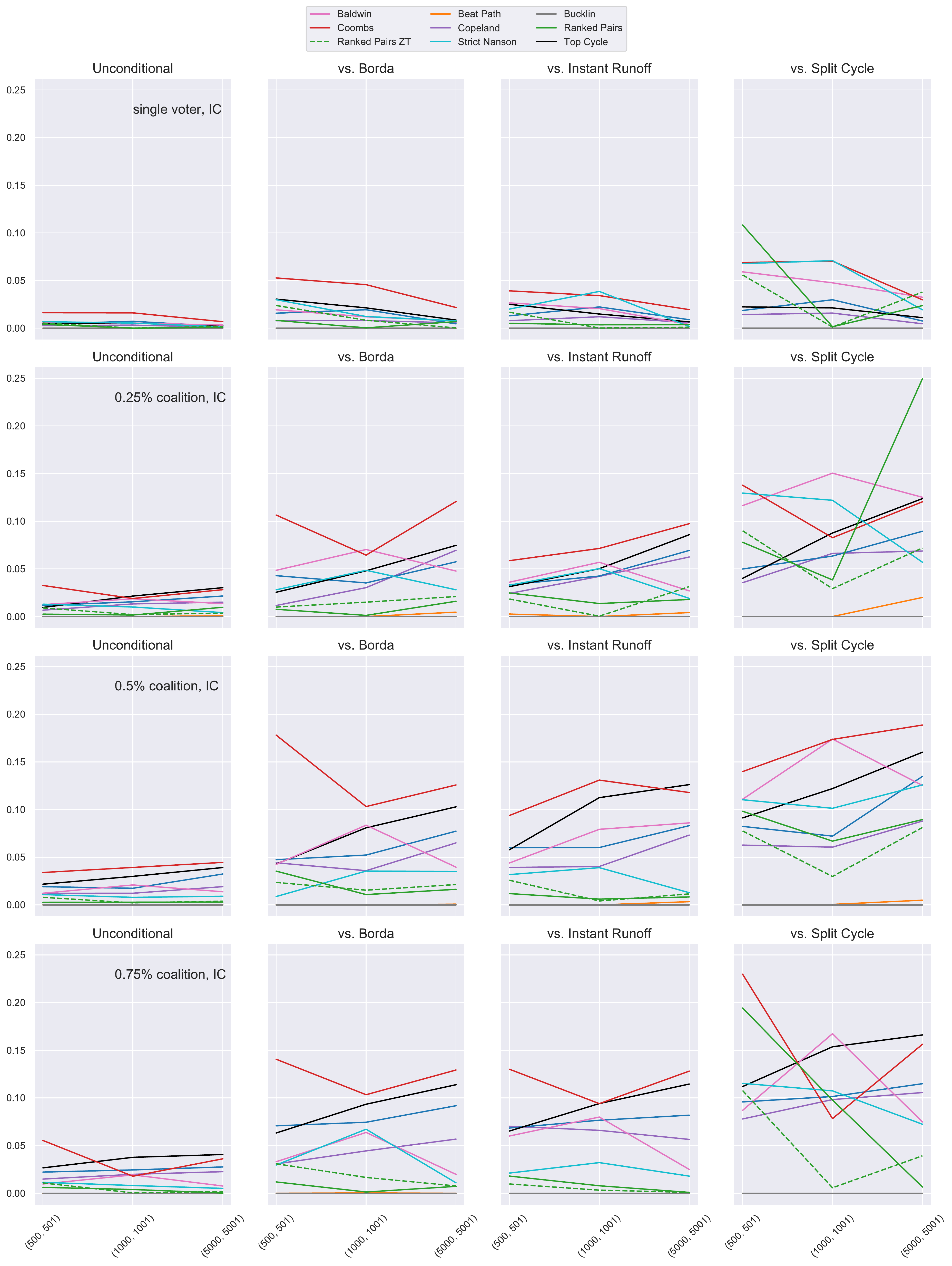}

\caption{probability of a profile-coalition pair $(\mathbf{P},\mathbf{P}_C)$ witnessing a violation of PI either unconditionally (far left), conditional on disagreement  in $\mathbf{P}$ or $\mathbf{P}+\mathbf{P}_C$  with    Borda (second from left),  with Instant Runoff (second from right), or with Split Cycle (far right). The first row is for a single voter coalition, the second for coalitions of 0.25\% of initial total voter size, the third for 0.5\%, and the fourth for 0.75\%.}\label{Absolute+ConditionalAddition}
\end{figure*}

As suggested by earlier results (Section \ref{ProfileCondition}), conditioning on the probability that in our random profile-coalition pair $(\mathbf{P},\mathbf{P}_C)$ the voting method $F_1$ disagrees with the method $F_2$ that satisfies PI---where disagreement now means that $F_1(\mathbf{P})\neq F_2(\mathbf{P})$ or $F_1(\mathbf{P}+\mathbf{P}_C)\neq F_2(\mathbf{P}+\mathbf{P}_C)$ (recall Remark \ref{ConditioningRemark})---dramatically increases the probability that $F_1$ violates PI. The results are shown in columns 2-4 of Figure \ref{Absolute+ConditionalAddition}.

\subsubsection{Probability of PI violation relativized to voter potency}\label{ProfileCoalitionRelativized}

We can also apply the idea of relativizing to voter potency from Section \ref{ProfileRelativize} in the paradigm of sampling profile-coalition pairs. In particular, we want to estimate the value of the following ratio for a random profile-coalition pair $(\mathbf{P},\mathbf{P}_C)$, a randomly chosen $a\in F(\mathbf{P})$, and a random new voter:
\begin{equation}\frac{Pr(\mbox{the new voters in $C$ cause $a$ to lose} \mid \mbox{$a$ is the favorite of the new voters in $C$})}{Pr(\mbox{the new voters in $C$ cause $a$ to lose})}.\label{Ratio2}\end{equation}
Intuitively, the numerator should be 0, but as we know, for voting methods that violate PI it is not zero. It is small, but on the other hand, the denominator is also small, since small coalitions of voters have limited influence in any election. Thus, as in Section \ref{ProfileRelativize}, the ratio itself can be surprisingly large. Ratios of the form $Pr(A\mid B)/Pr(A)$, as in (\ref{Ratio2}), are used in confirmation theory to measure the degree of support that evidence provides for a hypothesis (see, e.g., \cite[p.~54]{Horwich2016}). Note that $Pr(A\mid B)/Pr(A)=Pr(B\mid A)/Pr(B)$. Thus, we can phrase our question in one of two ways: If you learn that the new voters in $C$ rank $a$ first, to what extent does this support the hypothesis that the voters in $C$ will cause $a$ to lose? Or, alternatively, if you learn that the new voters in $C$ caused $a$ to lose, to what extent does this support the hypothesis that the voters in $C$ ranked $a$ first? Of course for methods satisfying PI the answer is ``not at all.''

\begin{figure*}[!htbp]

\centering
\includegraphics[scale=0.4]{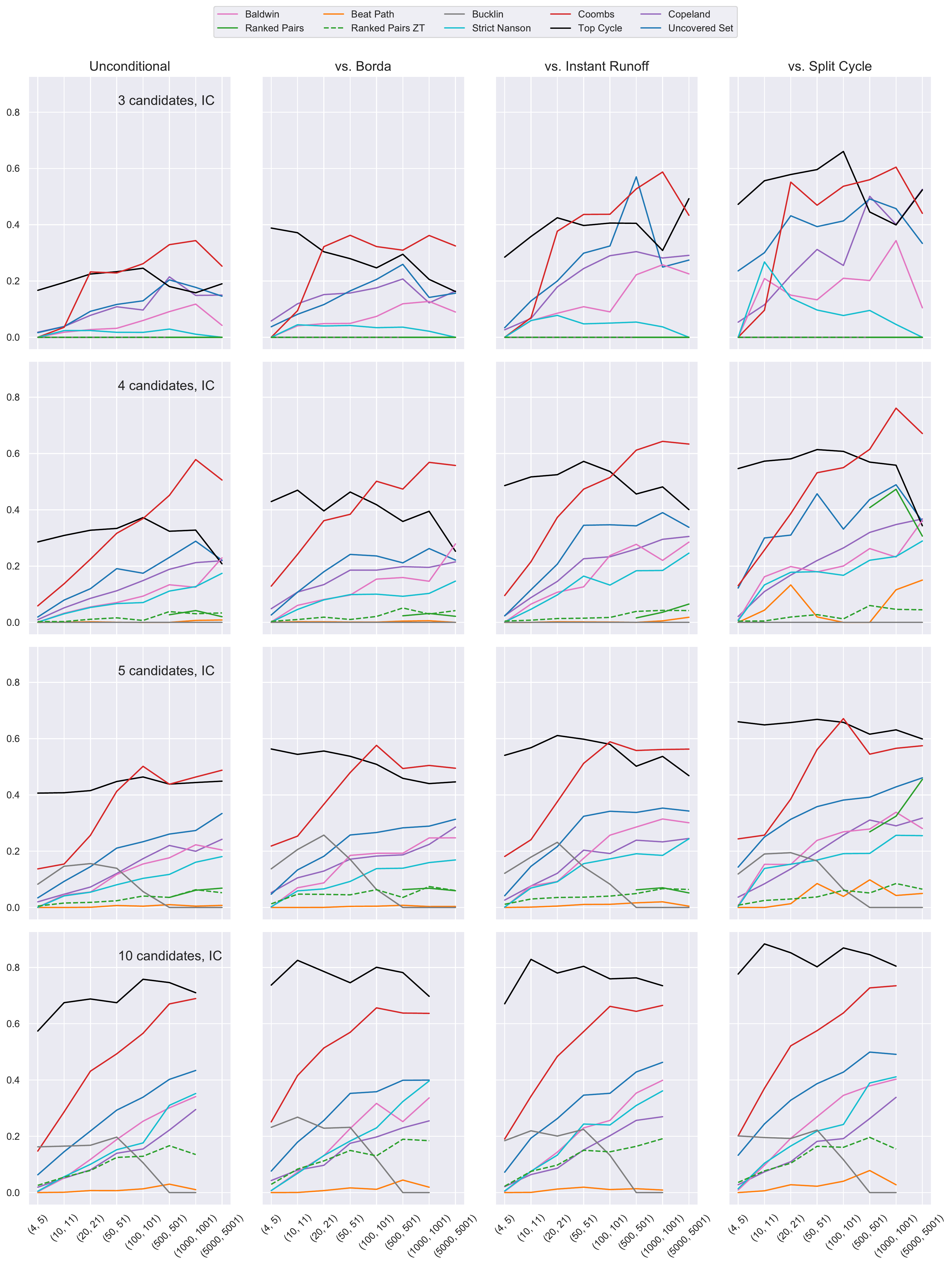}

\caption{the probability ratio in (\ref{Ratio2}) from Section \ref{ProfileCoalitionRelativized} for profile-coalition pairs $(\mathbf{P},\mathbf{P}_C)$ with $|C|=1$, either unconditionally (far left), conditional on disagreement with Borda in $\mathbf{P}$ or $\mathbf{P}+\mathbf{P}_C$ (second from left), conditional on disagreement with Instant Runoff in $\mathbf{P}$ or $\mathbf{P}+\mathbf{P}_C$, or conditional on disagreement with Split Cycle in $\mathbf{P}$ or $\mathbf{P}+\mathbf{P}_C$ (far right). Due to computational limitations, results for 10 candidates and 5,000/5,001 voters were not obtained, and results for Ranked Pairs were obtained only for 3-5 candidates and 500-5,001 voters (as increasing the number of voters decreases the likelihood of tied margins).}\label{Absolute+ConditionalRelativizedAddition}
\end{figure*}

  Results are shown in Figure \ref{Absolute+ConditionalRelativizedAddition}, where columns 2-4 also condition the  probabilities in (\ref{Ratio2}) on $F_1$ disagreeing with $F_2$ in $\mathbf{P}$ or $\mathbf{P}+\mathbf{P}_C$. Note the strikingly high ratios for some methods---especially Coombs and Top Cycle, but also Uncovered Set, Baldwin, Nanson, and Copeland---as we increase the candidates.

\subsection{Other probability models}\label{OtherProbModels}

In addition to sampling profiles with the IC model, we tried several other probability models. In the P\'{o}lya-Eggenberger urn model \cite{Berg1985}, each voter in turn randomly draws a linear order from an urn. Initially the urn is $\mathcal{L}(X)$. If a voter randomly chooses $L$ from the urn, we return $L$ to the urn plus $\alpha\in\mathbb{N}$ copies of $L$. IC is the special case where $\alpha=0$. The Impartial Anonymous Culture (IAC) is the special case where $\alpha=1$. We also considered $\alpha=10$ (as in \cite{Brandt2014,Brandt2019,Brandt2020}) for the model we call URN. In the Mallow's model (see \cite{Mallow1957,Marden1995}), given a reference ranking $L_0\in\mathcal{L}(X)$ and $\phi\in (0,1]$, the probability that a voter's ballot is $L\in\mathcal{L}(X)$ is $Pr_{L_0}(L)=\phi^{\tau(L,L_0)}/C$ where $\tau(L,L_0)= {{|X|}\choose{2}} - |L\cap L_0|$, the Kendell-tau distance of $L$ to $L_0$, and $C$ is a normalization constant. We  considered two reference rankings, $L_0$ and its converse $L_0^{-1}$ (e.g., $L_0$ ranks candidates from more liberal to more conservative, and $L_0^{-1}$ vice versa), in which case the probability that a voter's ballot is $L$ is $\frac{1}{2} Pr_{L_0}(L)+\frac{1}{2}Pr_{L_0^{-1}}(L)$. We set $\phi=0.8$ (as in \cite{Brandt2019,Brandt2020}).

\begin{figure*}[!htbp]
\centering
\includegraphics[scale=0.4]{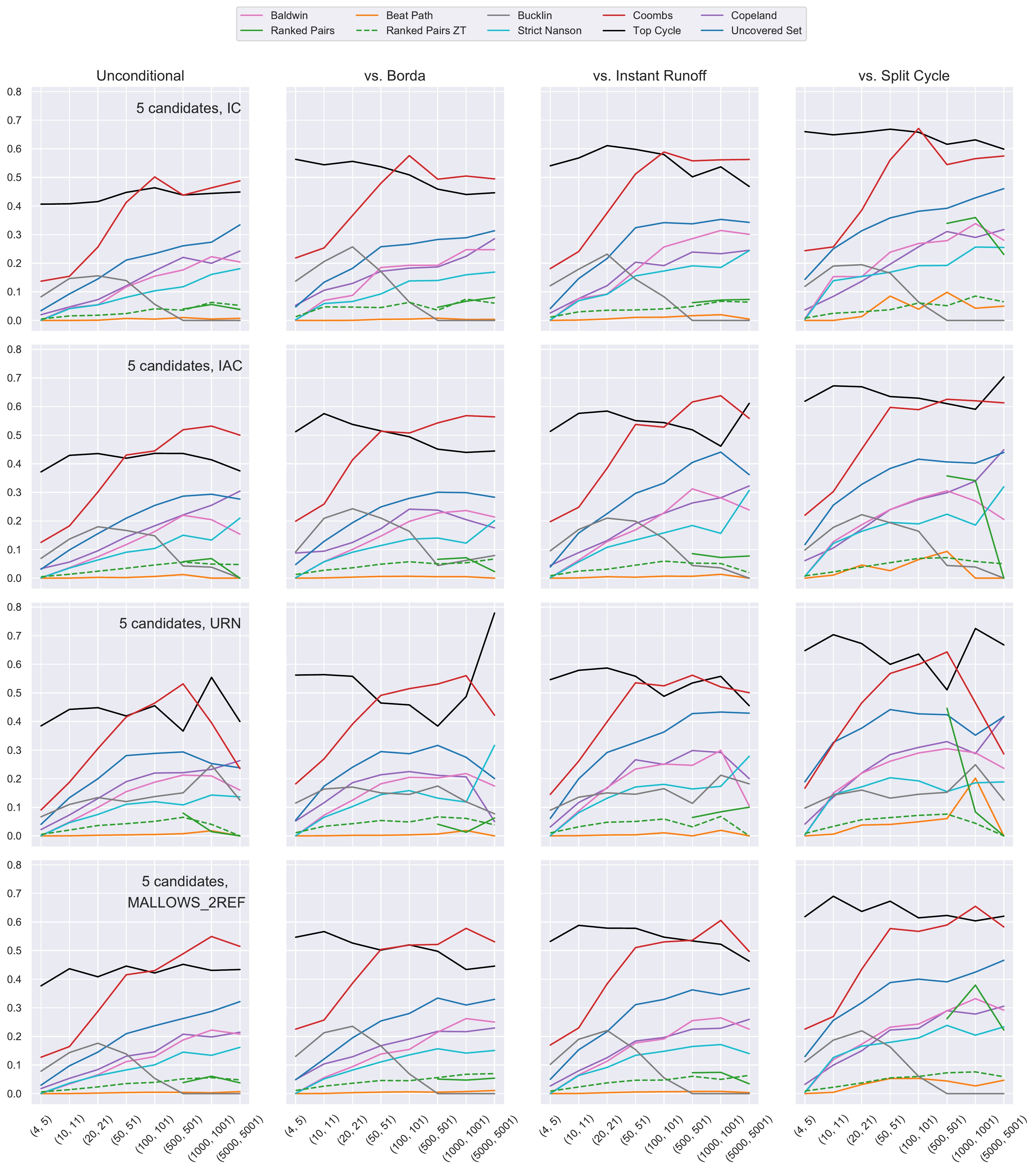}

\caption{the same measures as in row 3 of Figure \ref{Absolute+ConditionalRelativizedAddition} but under different probability models. As before, due to computational limitations, results for Ranked Pairs were obtained only for 500-5,001 voters.}\label{DiffModels}
\end{figure*}

The most important finding concerning the different probability models is that as we deviate from the IC model, the main phenomena seen in the previous simulations do not disappear. Figure \ref{DiffModels} shows the 5-candidate case of Figure \ref{Absolute+ConditionalRelativizedAddition} under these three probability models, plus IC again (for easy comparison). We see that the surprisingly high values for the ratio measure in (\ref{Ratio2}) are not artifacts of the IC model. 

\section{Conclusion}\label{Conclusion}

In their book on failures of monotonicity-type axioms for voting methods, Nurmi and Felsenthal \cite{Felsenthal2017} argue that violations of positive (and negative) involvement are the ``more dramatic types of no show paradox,'' wherein ``the issues of legitimacy of outcomes and questionable voter incentives are far more obvious'' (p.~86). In this paper, we proposed ways of measuring the extent to which these dramatic violations occur. The next step in future work is to marshal these measures when arguing in favor of some voting methods over others. Under certain voting methods that violate PI, much of a voter's  potency is turned against them---in particular, against their desire to see their favorite candidate elected. It is like adding insult to injury to be told that not only does your vote have little chance of influencing the election, but also your vote may cause your \textit{favorite} candidate to lose, and \textit{the probability of your vote doing so is not at all insignificant compared to the probability of it causing any candidate to lose}. The probabilities may be large enough to raise real concerns in small elections in committees and clubs, unless one somehow knows in advance that none of the probability models used here is relevant for the committee (e.g., one knows in advance that the committee has single-peaked preferences). But for large elections,  defenders of voting methods that violate PI may simply respond: don't worry---your vote probably won't cause your favorite to lose, because it probably won't influence the election at all. The same response could be used to try to dismiss concerns about violations of other single-voter axioms, such as monotonicity or single-voter strategyproofness. Such responses raise what is perhaps the ultimate paradox of voting: why do voters vote in large elections? The significance of single-voter axioms for large elections may turn on the answer to that question.

\appendix

\section{Appendix}\label{Appendix}

The following are examples of profiles witnessing PI violation for the voting methods from Proposition~\ref{MethodsViolatePI}. In the profiles, the new voter's ranking is highlighted in grey.\medskip  

\begin{flushleft}
\begin{tabular}{clc}
\begin{minipage}[t]{3in}
\begin{example}[Baldwin violates PI]  Candidate $c$ is the winner on the left (the order of elimination is $a$, $d$, $b$) but not on the right (the order of elimination is $d$, $b$, $c$):  \medskip

\begin{minipage}{1.35in}
\begin{center}
\begin{tabular}{ccccc}

$1$ & $2$ & $1$ & $1$ & $1$    \\\hline
$b$ & $c$ &  $d$ & $d$ & $a$   \\
$a$ & $b$ &  $a$ & $b$ & $c$ \\
$d$ & $d$ &  $c$ & $a$ & $b$ \\
$c$ & $a$ &  $b$ & $c$ & $d$ \\
\end{tabular}\\[5pt]
Baldwin winners: $\{c\}$
\end{center}
\end{minipage}\begin{minipage}{1.7in}
\begin{center}
\begin{tabular}{cccccc}
$1$ & $2$ & $1$ & $1$ & $1$ & $1$   \\\hline
$b$ & $c$ &  $d$ & $d$ & $a$ & {\cellcolor[gray]{.8}}${c}$   \\
$a$ & $b$ &  $a$ & $b$ & $c$ & {\cellcolor[gray]{.8}}${a}$ \\
$d$ & $d$ &  $c$ & $a$ & $b$ & {\cellcolor[gray]{.8}}${b}$ \\
$c$ & $a$ &  $b$ & $c$ & $d$ & {\cellcolor[gray]{.8}}${d}$ \\
\end{tabular}\\[5pt]
Baldwin winners: $\{a\}$
\end{center}
\end{minipage}\medskip

\noindent Since there are no ties for the lowest Borda score at any stage, the same holds for Baldwin PUT.
\end{example}
\end{minipage} & 

\begin{minipage}[t]{3in}
\begin{example}[Beat Path violates PI]  The left graph below can be realized as the margin graph of a profile with 11 voters;  the right graph is obtained by adding a voter with the ranking $b\ a\ c\ d$.  Then $b$ is a winner on the left but not on the right:

\begin{minipage}{1.6in}
\begin{tikzpicture}
\node (a) at (4,1.5) {$a$};
\node (b) at (2.5, 0) {$b$};
\node (d) at (5.5,0) {$d$};
\node (c) at (4,-1.5) {$c$};
\path[->, draw, thick] (a) to node[fill=white] {\small $1$} (b);
\path[->, draw, thick] (a) to node[fill=white] {\small $1$} (d);
\path[->, draw, thick] (b) to node[fill=white, pos=0.7] {\small $3$} (d);
\path[->, draw, thick] (c) to node[fill=white] {\small $3$} (b);
\path[->, draw, thick] (c) to node[fill=white, pos=0.7] {\small $1$} (a);
\path[->, draw, thick] (d) to node[fill=white] {\small $3$} (c);

\node at (4,-2.25) {Beat Path: $\{a, b, c, d\}$};
\end{tikzpicture}\end{minipage}\begin{minipage}{1.5in}\begin{tikzpicture}
\node (a) at (4,1.5) {$a$};
\node (b) at (2.5, 0) {$b$};
\node (d) at (5.5,0) {$d$};
\node (c) at (4,-1.5) {$c$};
\path[->, draw, thick] (a) to node[fill=white] {\small $2$} (d);
\path[->, draw, thick] (b) to node[fill=white, pos=0.5] {\small $4$} (d);
\path[->, draw, thick] (c) to node[fill=white] {\small $2$} (b);
\path[->, draw, thick] (d) to node[fill=white] {\small $2$} (c);

\node at (4,-2.25) {Beat Path: $\{a\}$};
\node at (4,-2.75) {New ranking: $b\ a\ c\ d$};
\end{tikzpicture}
\end{minipage}
\end{example}
\end{minipage}
\end{tabular}
\end{flushleft}

\begin{flushleft}
\begin{tabular}{clc}
\begin{minipage}[t]{3in}
\begin{example}[Bucklin violates PI] Candidate $c$ is the winner on the left (there are no 1st or 2nd level majority winners;  $a$ and $c$ are 3rd level majority winners; and more voters rank $c$ in 3rd place or higher than $a$) but not on the right (because $e$ is the unique 2nd level majority winner):\medskip 

\begin{minipage}{1.4in}
\begin{center}
\begin{tabular}{cccc}

$1$ & $1$ & $1$ & $1$   \\\hline
$a$ & $a$ &  $b$ & $c$   \\
$b$ & $e$ &  $e$ & $d$  \\
$c$ & $c$ &  $c$ & $a$   \\
$e$ & $b$ &  $d$ & $b$   \\
$d$ & $d$ &  $a$ & $e$   \\
\end{tabular}\\[5pt]
Bucklin: $\{c\}$ 
\end{center}
\end{minipage}\begin{minipage}{1.5in}
\begin{center}
\begin{tabular}{ccccc}
$1$ & $1$ & $1$ & $1$  & $1$  \\\hline
$a$ & $a$ &  $b$ & $c$ &  {\cellcolor[gray]{.8}}${c}$  \\
$b$ & $e$ &  $e$ & $d$  &  {\cellcolor[gray]{.8}}${e}$ \\
$c$ & $c$ &  $c$ & $a$  &  {\cellcolor[gray]{.8}}${b}$  \\
$e$ & $b$ &  $d$ & $b$ &  {\cellcolor[gray]{.8}}${d}$   \\
$d$ & $d$ &  $a$ & $e$  &  {\cellcolor[gray]{.8}}${a}$  \\

\end{tabular}\\[5pt]
Bucklin: $\{e\}$
\end{center}
\end{minipage}\medskip

The same example works for Simplified Bucklin, as defined in Footnote \ref{SimplifiedBucklin}, only the winners on the left are $a$ and $c$, and the winner on the right is $e$.
\end{example}

\end{minipage} & 

\begin{minipage}[t]{3in}
\begin{example}[Coombs violates PI] Candidate $a$ is the winner on the left (because $b$ and $c$ are eliminated in the first round, and then $a$ is the majority winner) but not on the right (because $b$ is eliminated first, and then $c$ is the majority winner): \medskip

\begin{minipage}{1.3in}
\begin{center}
\begin{tabular}{ccccc}
$1$ & $2$ & $1$  & $1$  & $1$  \\\hline
$a$ & $c$ &  $a$ &  $b$ & $c$  \\
$b$ & $a$ &  $d$ & $c$ & $b$  \\
$d$ & $d$ &  $b$ & $d$ & $a$ \\
$c$ & $b$ &  $c$ & $a$ & $d$ \\
\end{tabular}\\[5pt]
Coombs: $\{a\}$
\end{center}
\end{minipage}\begin{minipage}{1.7in}
\begin{center}
\begin{tabular}{cccccc}
$1$ & $2$ & $1$  & $1$  & $1$ & $1$  \\\hline
$a$ & $c$ &  $a$ &  $b$ & $c$  &{\cellcolor[gray]{.8}}${a}$  \\
$b$ & $a$ &  $d$ & $c$ & $b$  &{\cellcolor[gray]{.8}}${d}$  \\
$d$ & $d$ &  $b$ & $d$ & $a$  &{\cellcolor[gray]{.8}}${c}$ \\
$c$ & $b$ &  $c$ & $a$ & $d$  &{\cellcolor[gray]{.8}}${b}$ \\
\end{tabular}\\[5pt]
Coombs: $\{c\}$\\
\end{center}
\end{minipage}\medskip

\noindent Under Coombs PUT, on the left, since $b$ and $c$ tie for most last place votes, we consider two cases: (i) if we eliminate $b$ in the first round, then $c$ is the majority winner; (ii) if we eliminate $c$ in the first round, then $a$ is the majority winner. Hence both $a$ and $c$ win in the profile on the left. In the profile on the right, $c$ is the unique winner for Coombs PUT for the same reason as for Coombs.
\end{example}\end{minipage}
\end{tabular}
\end{flushleft}\medskip

\begin{example}[Copeland and Uncovered Set violate PI]\label{CopelandExample}  Candidate $b$ is a Copeland winner on the left but not on the right:\footnote{The same example provides a PI violation for the Llull method defined in Footnote \ref{LlullNote}. The Llull winners are the same as the Copeland winners in both profiles.}
\begin{center}
\begin{minipage}{2in}
\begin{center}
\begin{tikzpicture}
\node at (-2,0) {\begin{minipage}{2.5in}\begin{center}\begin{tabular}{cccc}
$2$ & $1$ & $2$    \\\hline
$a$ & $b$ &  $c$   \\
$b$ & $c$ &  $a$  \\
$c$ & $a$ &  $b$  \\
\end{tabular}\\[5pt]
\end{center}
\end{minipage}};

\node (a) at (1,0.75) {$a$};
\node (b) at (0,-0.75) {$b$};
\node (c) at (2,-0.75) {$c$};
\path[->, draw, thick] (a) to node[fill=white] {\small $3$} (b);
\path[->, draw, thick] (b) to node[fill=white] {\small $1$} (c);
\path[->, draw, thick] (c) to node[fill=white] {\small $1$} (a);

\node at (-0.25,-1.5) {Copeland winners: $\{a,b,c\}$};
\node at (-0.25,-2) {Gillies Uncovered Set winners: $\{a,b,c\}$};
\node at (-0.25,-2.5) {Fishburn Uncovered Set winners: $\{a,b,c\}$};
\end{tikzpicture}
\end{center}
\end{minipage}\qquad\qquad\quad\begin{minipage}{3.5in}
\begin{center}
\begin{tikzpicture}
\node at (-2,0) {\begin{minipage}{2.5in}\begin{center}\begin{tabular}{cccc}
$2$ & $1$ & $2$  & 1  \\\hline
$a$ & $b$ &  $c$ &{\cellcolor[gray]{.8}}${b}$  \\
$b$ & $c$ &  $a$ &{\cellcolor[gray]{.8}}${a}$ \\
$c$ & $a$ &  $b$ &{\cellcolor[gray]{.8}}${c}$ \\
\end{tabular}\\[5pt]
\end{center}
\end{minipage}};

\node (a) at (1,0.75) {$a$};
\node (b) at (0,-0.75) {$b$};
\node (c) at (2,-0.75) {$c$};
\path[->, draw, thick] (a) to node[fill=white] {\small $2$} (b);
\path[->, draw, thick] (b) to node[fill=white] {\small $2$} (c);

\node at (-0.5,-1.5) {Copeland winners: $\{a\}$};
\node at (-0.5,-2) {Gillies Uncovered Set winners: $\{a,c\}$};
\node at (-0.5,-2.5) {Fishburn Uncovered Set winners: $\{a\}$};
\end{tikzpicture}
\end{center}
\end{minipage}
\end{center}
The same example works for the Gillies (resp.~Fishburn) version of Uncovered Set (recall Remark \ref{TiesRemark}): on the left, the winners are $a$, $b$, and $c$, while on the right, the winners are $a$ and $c$ (resp.~$a$). \end{example}

\begin{example}[Strict Nanson violates PI] Candidate $c$ is the winner on the left (the order of elimination is $b$, $d$, $a$) but not on the right (both $a$ and $b$ are eliminated in the first round, followed by $c$):\medskip

\begin{minipage}{2.5in}
\begin{center}
\begin{tabular}{cccccc}

$1$ & $1$ & $3$ & $3$ & $1$ & $1$ \\\hline
$a$ & $d$ &  $c$ & $a$ & $d$ & $d$   \\
$c$ & $b$ &  $b$ & $d$ & $c$ & $c$ \\
$d$ & $c$ &  $a$ & $c$ & $a$ & $b$ \\
$b$ & $a$ &  $d$ & $b$ & $b$ & $a$ \\
\end{tabular}\\[5pt]
Strict Nanson winners: $\{c\}$
\end{center}
\end{minipage}\qquad\begin{minipage}{2.5in}
\begin{center}
\begin{tabular}{ccccccc}
$1$ & $1$ & $3$ & $3$ & $1$ & $1$ & $1$ \\\hline
$a$ & $d$ &  $c$ & $a$ & $d$ & $d$ & {\cellcolor[gray]{.8}}${c}$   \\
$c$ & $b$ &  $b$ & $d$ & $c$ & $c$ & {\cellcolor[gray]{.8}}${d}$ \\
$d$ & $c$ &  $a$ & $c$ & $a$ & $b$ & {\cellcolor[gray]{.8}}${b}$ \\
$b$ & $a$ &  $d$ & $b$ & $b$ & $a$ & {\cellcolor[gray]{.8}}${a}$ \\
\end{tabular}\\[5pt]
Strict Nanson winners: $\{d\}$
\end{center}
\end{minipage}
\end{example}

\begin{example}[Ranked Pairs violates PI]  The left graph below can be realized as the margin graph of a profile with 20 voters; then the right graph below is obtained by adding a voter with the ranking $d\ a\ b\ c$.  Then candidate $d$ is a  winner on the left (we first lock in the edges $(a,c)$, $(d,c)$, and $(d,b)$, and then there is a choice of whether to prioritize the $(b,a)$ edge, in which case $d$ wins, or the $(a,d)$ edge, in which case $a$ wins) but not on the right (we first lock in the edges $(a,c)$, $(d,c)$, and $(d,b)$, and then there is a choice of whether to prioritize $(c,b)$ or $(a,d)$, but in either case $(a,d)$ eventually gets locked in):

\begin{center}
\begin{minipage}{2.5in}
\begin{tikzpicture} 
\node (a) at (3.5,1.5) {$a$};
\node (b) at (2, 0) {$b$};
\node (d) at (5,0) {$d$};
\node (c) at (3.5,-1.5) {$c$};
\path[->, draw, thick] (a) to node[fill=white, pos=0.7] {\small $8$} (c);
\path[->, draw, thick] (a) to node[fill=white, pos=0.5] {\small $2$} (d);
\path[->, draw, thick] (b) to node[fill=white] {\small $2$} (a);
\path[->, draw, thick] (c) to node[fill=white] {\small $2$} (b);
\path[->, draw, thick] (d) to node[fill=white, pos=0.7] {\small $4$} (b);
\path[->, draw, thick] (d) to node[fill=white] {\small $6$} (c);

\node at (3.5,-2) {Ranked Pairs winners: $\{a, d\}$};
\node at (3.5, -2.5) {\textcolor{white}{New ranking: $d\ a\ c\ b$}};
\end{tikzpicture}\end{minipage}\qquad \begin{minipage}{2.5in}\begin{tikzpicture} 
\node (a) at (3.5,1.5) {$a$};
\node (b) at (2, 0) {$b$};
\node (d) at (5,0) {$d$};
\node (c) at (3.5,-1.5) {$c$};
\path[->, draw, thick] (a) to node[fill=white, pos=0.7] {\small $9$} (c);
\path[->, draw, thick] (a) to node[fill=white, pos=0.5] {\small $1$} (d);
\path[->, draw, thick] (b) to node[fill=white] {\small $1$} (a);
\path[->, draw, thick] (c) to node[fill=white] {\small $3$} (b);
\path[->, draw, thick] (d) to node[fill=white, pos=0.7] {\small $5$} (b);
\path[->, draw, thick] (d) to node[fill=white] {\small $7$} (c);

\node at (3.5,-2) {Ranked Pairs winners: $\{a\}$};
\node at (3.5, -2.5) {New ranking: $d\ a\ c\ b$};
\end{tikzpicture}
\end{minipage}
\end{center}
For Ranked Pairs ZT, in the profile on the left below, where the tiebreaking voter's ranking is indicated with $(\star)$, $b$ is the  winner (we lock in $(d,c)$, $(b,c)$, $(b,d)$, and then $(c,a)$). But on the right, $a$ is the winner (there are no cycles, so all edges are locked in, and then the ranking $(\star)$ ranks $a$ above $b$).

\begin{center}
\begin{tikzpicture} 
\node at (0,0) {\begin{minipage}{1in}\begin{center}\begin{tabular}{cccc}
$1$ $(\star)$ & $2$ & $2$    \\\hline
$c$ & $a$ &  $b$   \\
$a$ & $b$ &  $d$  \\
$d$ & $d$ &  $c$  \\
$b$ & $c$ &  $a$  \\
\end{tabular}\\[5pt]
\end{center}
\end{minipage}};

\node (a) at (3.5,1.5) {$a$};
\node (b) at (2, 0) {$b$};
\node (d) at (5,0) {$d$};
\node (c) at (3.5,-1.5) {$c$};
\path[->, draw, thick] (a) to node[fill=white, pos=0.5] {\small $1$} (b);
\path[->, draw, thick] (a) to node[fill=white, pos=0.5] {\small $1$} (d);
\path[->, draw, thick] (b) to node[fill=white] {\small $3$} (c);
\path[->, draw, thick] (b) to node[fill=white, pos=0.7] {\small $3$} (d);
\path[->, draw, thick] (c) to node[fill=white, pos=0.7] {\small $1$} (a);
\path[->, draw, thick] (d) to node[fill=white, pos=0.5] {\small $3$} (c);

\node at (3.5,-2) {Ranked Pairs ZT winners: $\{b\}$};
\end{tikzpicture}\quad\begin{tikzpicture} 

\node at (-.8,0) {\begin{minipage}{1in}\begin{center}\begin{tabular}{cccc}
$1$ $(\star)$ & $2$ & $2$ & $1$    \\\hline
$c$ & $a$ &  $b$ & {\cellcolor[gray]{.8}}${b}$   \\
$a$ & $b$ &  $d$ & {\cellcolor[gray]{.8}}${a}$  \\
$d$ & $d$ &  $c$ & {\cellcolor[gray]{.8}}${c}$  \\
$b$ & $c$ &  $a$ & {\cellcolor[gray]{.8}}${d}$  \\
\end{tabular}\\[5pt]
\end{center}
\end{minipage}};

\node (a) at (3.5,1.5) {$a$};
\node (b) at (2, 0) {$b$};
\node (d) at (5,0) {$d$};
\node (c) at (3.5,-1.5) {$c$};
\path[->, draw, thick] (a) to node[fill=white, pos=0.5] {\small $2$} (d);
\path[->, draw, thick] (b) to node[fill=white] {\small $4$} (c);
\path[->, draw, thick] (b) to node[fill=white, pos=0.7] {\small $4$} (d);
\path[->, draw, thick] (d) to node[fill=white, pos=0.5] {\small $2$} (c);

\node at (3.5,-2) {Ranked Pairs ZT winners: $\{a\}$};
\end{tikzpicture}
\end{center}
\end{example}

\begin{example}[Top Cycle violates PI] Candidate $b$ is a winner on the left but not the right:
\begin{center}
\begin{minipage}{2in}
\begin{center}
\begin{tikzpicture}
\node at (-2,0) {\begin{minipage}{2.5in}\begin{center}\begin{tabular}{cccc}
$1$ & $1$      \\\hline
$a$ & $c$   \\
$b$ & $a$   \\
$c$ & $b$   \\
\end{tabular}\\[5pt]
\end{center}
\end{minipage}};

\node (a) at (1,0.75) {$a$};
\node (b) at (0,-0.75) {$b$};
\node (c) at (2,-0.75) {$c$};
\path[->, draw, thick] (a) to node[fill=white] {\small $2$} (b);

\node at (-0.25,-1.5) {Top Cycle winners: $\{a,b,c\}$};
\end{tikzpicture}
\end{center}
\end{minipage}\qquad\qquad\begin{minipage}{3.5in}
\begin{center}
\begin{tikzpicture}
\node at (-2,0) {\begin{minipage}{2.5in}\begin{center}\begin{tabular}{ccc}
$1$ & $1$ &  $1$  \\\hline
$a$ & $c$ &   {\cellcolor[gray]{.8}}${b}$  \\
$b$ & $a$ &  {\cellcolor[gray]{.8}}${a}$ \\
$c$ & $b$ & {\cellcolor[gray]{.8}}${c}$ \\
\end{tabular}\\[5pt]
\end{center}
\end{minipage}};

\node (a) at (1,0.75) {$a$};
\node (b) at (0,-0.75) {$b$};
\node (c) at (2,-0.75) {$c$};
\path[->, draw, thick] (a) to node[fill=white] {\small $1$} (b);
\path[->, draw, thick] (a) to node[fill=white] {\small $1$} (c);
\path[->, draw, thick] (b) to node[fill=white] {\small $1$} (c);

\node at (-0.5,-1.5) {Top Cycle winners: $\{a\}$};
\end{tikzpicture}
\end{center}
\end{minipage}
\end{center}
\end{example}

\bibliographystyle{eptcs}
\bibliography{PI}
\end{document}